\pdfoutput=1
\documentclass{rnaastex}

\begin{document}

\title{The Recurrent Nova Candidate M31N 1966-08a = 1968-10c is a Galactic Flare Star}

\email{Email: ashafter@sdsu.edu,
 mhenze@sdsu.edu,
 m.j.darnley@ljmu.ac.uk,
 rbc@astro.psu.edu,
 bdd8@psu.edu,
 slhawley@uw.edu}

\author{A. W. Shafter}
\altaffiliation{Department of Astronomy, San Diego State University}

\author{M. Henze}
\altaffiliation{Department of Astronomy, San Diego State University}

\author{M. J. Darnley}
\altaffiliation{Astrophysics Research Institute, Liverpool John Moores University}

\author{R. Ciardullo}
\altaffiliation{Department of Astronomy, Pennsylvania State University}

\author{B. D. Davis}
\altaffiliation{Department of Astronomy, Pennsylvania State University}

\author{S. L. Hawley}
\altaffiliation{Department of Astronomy, University of Washington}




\keywords{galaxies: individual (M31); galaxies: stellar content; stars: individual (M31N 1966-08a); stars: flare; novae, cataclysmic variables}

\section{} 

\citet{ros64,ros73} and \citet{ros89} conducted a multi-decade (1955 -- 1986)
imaging survey for novae in M31
using the 1.22m reflector at the Asiago observatory, supplemented
from 1973 onwards with observations from the 1.82m telescope at Mount Ekar.
During the course
of the survey, Rosino discovered a total of 142 nova candidates in M31.
As described in \citet{sha15}, based on spatial coincidence, a total
of six of these outbursts (R029 = M31N 1960-12a, R039 = 1962-11a,
R048 = 1963-09c, R066 = 1966-08a, R079 = 1968-09a, R081 = 1968-10c)
were found to be associated with 4 recurrent nova
candidates (M31N 1960-12a = 2013-05b, M31N 1926-06a = 1962-11a,
M31N 1963-09c = 1968-09a = 2010-10e = 2015-10c, M31N 1966-08a = 1968-10c).

Perhaps the most interesting of these eruptions is the
last pair, M31N 1966-08a = 1968-10c (R066 = R081),
which were observed on 12 August 1966 and 25 October 1968, respectively.
Available observations restrict the duration of the eruptions to less than
2 days. Not only is the
interval between eruptions ($\sim2.2$ yr) extremely short -- shorter
than any known recurrent nova with the exception of the remarkable
system M31N 2008-12a with a $\sim1$~yr recurrence time
\citep[e.g., see][and references therein]{dar17a, dar17b, hen17} -- but unlike
2008-12a (where an eruption has been detected every year for the past 9 years),
M31N 1966-08a was only seen in eruption twice in the past half century.
It would be surprising if M31N 1966-08a were a recurrent nova in M31 or
a foreground Galactic dwarf nova, as such systems typically remain
near maximum light for several days, or more. Given that
the object lies just 16$'$ from the center of M31 in a region of the
galaxy that is routinely monitored down to $m\sim18$ by a number of
amateur and professional astronomers alike, missing additional eruptions
would be particularly unlikely. On the other hand, the less predictable
and much shorter duration flares from dMe stars (hours)
could have more easily escaped detection.

To explore the possibility that M31N 1966-08a might be a foreground
Galactic flare star, we examined the deep photographic images of M31 from
\citet{mas06}. A stellar object, J004123.75+411459.6, with $V=20.3$ and
$U-B=1.9$, $V-R=1.5$, $R-I=2.0$ was found just
0.8$''$ south of the nominal position of M31N 1966-08a
({\it R.A.\/} = $00^{\mathrm h}~41^{\mathrm m}~23.75^{\mathrm s}$, {\it Decl.\/} = $+41^{\circ}~15'~00.4''$).
The star has also been detected in the infrared as
2MASS~00412375+4114596, with $J=15.3$, $H=14.9$, and $K=14.4$ \citep{sku06}.
Given its proximity, we have assumed that this star is
the progenitor of M31N 1966-08a and 1968-10c.
The implied outburst amplitude of $\sim3-4$
mag would be unusually low for a recurrent nova (without precedent,
in fact), but not unprecedented for a dMe star, which can have flares
of up to $5-6$ magnitudes
\citep[e.g., see][for the bright flares in AD Leo and YZ CMi, respectively]{haw91,kow10}.

To confirm the flare star hypothesis, on 2017 Nov. 23.27 UT
we obtained a spectrum of
J004123.75+411459.6 with the LRS2 spectrograph on the
Hobby-Eberly Telescope (HET). The spectrum, shown in Figure~1,
is that of a classic dMe flare star. The strength of the TiO bandheads
at $\sim$5430\AA\ and $\sim$6150\AA\
suggest a spectral type of approximately M3
\citep[see][]{boc07}, and the visibility of the
CaH feature at $\sim$6380\AA\ suggests that the star is a dwarf, not a giant.
Finally, we note that the measured emission-line
Balmer decrement (H$\alpha$:H$\beta$) is $\sim$4.2, which is
typical of dMe flare stars in quiescence. 

In summary, given the brevity of the two observed outbursts, the relatively
low outburst amplitude, and especially the spectrum
of the quiescent optical counterpart,
we conclude that M31N 1966-08a = 1968-10c is not a recurrent nova in M31
nor a Galactic dwarf nova,
but rather a Galactic dMe flare star projected
against the Andromeda galaxy.


\begin{figure}
\begin{center}
\includegraphics[scale=0.65,angle=0]{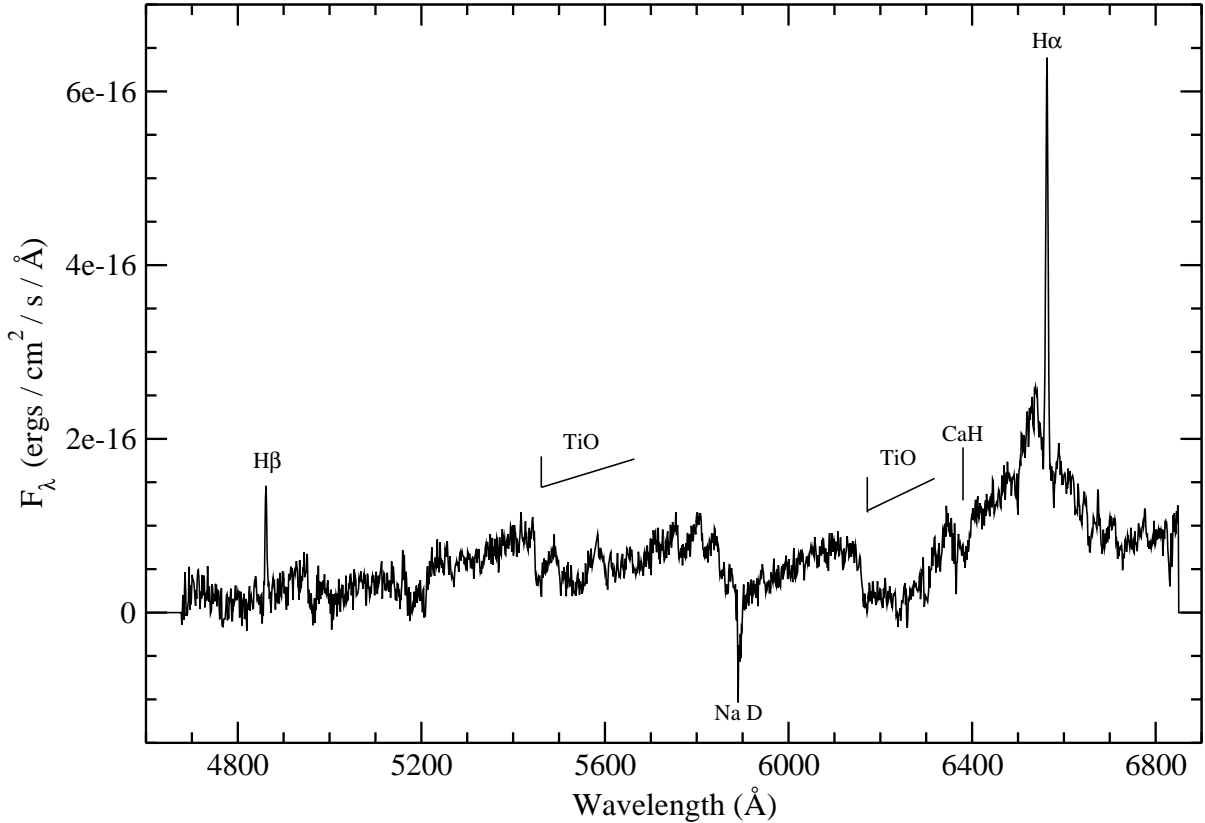}
\caption{The HET spectrum of the quiescent counterpart of M31N 1966-08a,
J004123.75+411459.6 (the absolute flux calibration is approximate).
The spectrum is typical of a dMe flare star, displaying
strong TiO bandheads at $\sim$5430\AA\ and $\sim$6150\AA,
along with narrow Balmer emission lines. Detection of the CaH absorption
feature at $\sim$6380\AA\ suggests that the star is a dwarf.
}
\end{center}
\end{figure}





\begin{thebibliography}{}

\bibitem[Bochanski et al.(2007)]{boc07} Bochanski, J. J., West, A. A., Hawley, S. L., \& Covey, K. R. 2007, AJ, 133, 531

\bibitem[Darnley et al.(2017a)]{dar17a} Darnley, M. J., Hounsell, R., Godon, P., Perley, D. A., Henze, M., et al. 2017a, ApJ, 847, 35  

\bibitem[Darnley et al.(2017b)]{dar17b} Darnley, M. J., Hounsell, R., Godon, P., Perley, D. A., Henze, M., et al. 2017b, ApJ, 849, 96

\bibitem[Hawley \& Pettersen(1991)]{haw91} Hawley, S. L., Pettersen, B. R. 1991,ApJ, 378, 725

\bibitem[Henze et al.(2017)]{hen17} Henze, M., et al. 2017, ApJ, submitted.

\bibitem[Kowalski et al.(2010)]{kow10} Kowalski, A. F., Hawley, S. L., Holtzman, J. A., Wisniewski, J. P., \& Hilton, E. J. 2010, ApJ, 714, 98

\bibitem[Massey et al.(2006)]{mas06} Massey, P., Olsen, K. A. G., Hodge, P. W., Strong, S. B., Jacoby, G. H., et al. 2006, AJ, 131, 2478

\bibitem[Rosino(1964)]{ros64} Rosino, L. 1964, Ann. Astrophys., 27, 498

\bibitem[Rosino(1973)]{ros73} Rosino, L. 1973, A\&AS, 9, 347 

\bibitem[Rosino et al.(1989)]{ros89} Rosino, L., Capaccioli, M., D'Onofrio, M., \& della Valle, M. 1989, AJ, 97, 83

\bibitem[Shafter et al.(2015)]{sha15} Shafter, A. W., Henze, M., Rector, T. A., Schweizer, F., Hornoch, K., et al. 2015, ApJS, 216, 34

\bibitem[Skrutskie et al.(2006)]{sku06} Skrutskie, M. F., Cutri, R. M., Stiening, R., Weinberg, M. D., Schneider, S., et al. 2006, AJ, 131, 1163


\end{thebibliography}
\end{document}